\documentclass[onecolumn,pra,groupedaddress,nobibnotes,showpacs,amssymb]{revtex4}

\usepackage{graphicx}
\usepackage{dcolumn}
\usepackage{bm}

\begin{document}

\title{Practical scheme for non-postselection entanglement concentration\\using linear optical elements}

\author{Myung-Joong Hwang}
\email{holy21@postech.ac.kr}
\affiliation{Department of Physics, Pohang University of Science and Technology (POSTECH), Pohang, 790-784, Korea}

\author{Yoon-Ho Kim}
\email{yoonho@postech.ac.kr}
\affiliation{Department of Physics, Pohang University of Science and Technology (POSTECH), Pohang, 790-784, Korea}

\date{To appear in Phys. Lett. A (2007), doi:10.1016/j.physleta.2007.04.098}

\begin{abstract}
We report a practical non-postselection entanglement concentration scheme in which a maximally entangled Bell-state photon pair is produced from two pairs of partially (or non-maximally) entangled photons.  Since this scheme is built only upon linear optical elements and does not require photon-number resolving detectors, it has immediate applications in experimental implementations of various quantum information protocols which require two-photon Bell-states.
\end{abstract}


\maketitle

\section{Introduction}

Quantum entanglement, typically defined as the non-local quantum correlation between the measured degrees of freedom of separate quantum particles, is now being recognized as the key element in the emerging field of quantum information. For photons, entanglement in polarization, among other photonic degrees of freedom, is particularly important in quantum information because the polarization state, i.e., the qubit, can be easily controlled with linear optical elements, such as, a half-wave plate or a quarter-wave plate. 

Although many protocols in quantum information are developed on the assumption that photons could be made into a specific maximally entangled state easily, preparing a maximally entangled state, a Bell-state, experimentally is not a trivial problem even in the case of two photons. To illustrate the problem more clearly, consider a Bell-state of two photons,  $|\Phi^{(+)}\rangle =\frac{1}{\sqrt{2}}(|H\rangle_1|H\rangle_2 + |V\rangle_1|V\rangle_2)$, where the subscripts refer to two separate photons. To prepare this state, it is necessary to ensure that the two two-photon probability amplitudes $|H\rangle_1|H\rangle_2$ and $|V\rangle_1|V\rangle_2$ exist with equal magnitudes, i.e., the normalized coefficients equal to $1/\sqrt{2}$. In addition, these amplitudes should be absent of any distinguishing information so that the amplitudes are coherently superposed with the fixed phase difference of 0 modulo $2\pi$. 

Experimentally, the coherent superposition between the two amplitudes could be tested by observing the quantum interference effects. Ensuring that the two amplitudes are of perfectly equal magnitudes, however, is a challenging experimental problem due to the practical issues (e.g., spatial and spectral mode matching) as well as many inherently fluctuating quantities (e.g., the detector dark counts, environmental noise, statistical fluctuations of the photon pair generation rate) that affect the joint detection rates. In practice, therefore, an experimentally prepared two-photon entangled state may be close to the ideal Bell-state but is often non-maximally entangled, even if all the distinguishing information is properly erased so that the state is pure.

In principle, entanglement concentration, i.e., the process of trading $N$ pairs of non-maximally entangled qubits for a qubit pair in a Bell-state, could be used to obtain a maximally entangled qubit pair \cite{Bennett}. In addition, it has been shown that entanglement swapping could aid the process of entanglement concentration \cite{Bose}.
In the recent years, a number of experimentally feasible linear optical entanglement concentration schemes for polarization (non-maximally) entangled photons have been proposed \cite{Yamamoto,Zhao} and demonstrated \cite{Yamamoto-exp,Zhao-exp}. For these schemes to really work, however, the projection measurements whose outcomes signal the successful entanglement concentration (i.e., preparation of a maximally entangled photon pair) must be performed with photon-number resolving detectors even in the case when the initial non-maximally entangled photon pairs are provided exactly as requested in the protocol. These schemes, therefore, rely on postselection of final four-photon detection events (amplitudes), thereby always destroying the desired output state itself.

Recently, a non-postselection entanglement concentration scheme using polarizing beamsplitters and waveplates was proposed \cite{Wang}. In the scheme of Ref.~\cite{Wang}, entanglement concentration without postselection could be achieved as long as the input state (two pairs of non-maximally entangled photons) is supplied as requested. 

In this paper, we propose a new non-postselection entanglement concentration scheme. The present scheme is quite practical as it requires only ordinary 50/50 and polarizing beamsplitters. The noteworthy feature of this scheme is that it works not only for the $|\Phi^{(+)}\rangle$-type non-maximally polarization-entangled photons, but also for the other three Bell-type non-maximally entangled states.

\begin{figure}[tbp]
\centering
\includegraphics[width=3.5in]{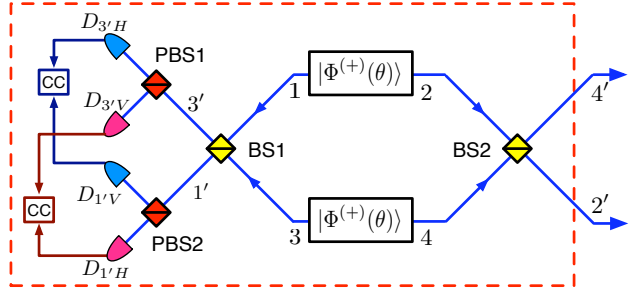}
\caption{Non-postselection entanglement concentration scheme using 50/50 (BS1 and BS2) and polarizing (PBS1 and PBS2) beamsplitters. Initially, the two photons pairs 1-2 and 3-4 are identically prepared in the same non-maximally entangled state. A count at the coincidence circuit (CC) between detectors $\{D_{1'H}, D_{3'V}\}$ or $\{D_{1'V},
D_{3'H}\}$ signals successful preparation of photons $2'$ and $4'$
in the singlet Bell-state $|\Psi^{(-)}\rangle_{2'4'}$. An interesting feature of this scheme is that the initial state of the two photon pairs can be any one of the four non-maximally entangled states, including $|\Phi^{(+)}(\theta)\rangle$. In addition, PBS1 and PBS2 need not necessarily be used in the H-V basis. See text for details. }\label{fig}
\end{figure}

\section{Non-postselection entanglement concentration scheme}

Let us now consider the non-postselection entanglement concentration scheme shown in Fig.~\ref{fig}. Photon pairs 1-2 and 3-4 are identically prepared in the non-maximally entangled state $|\Phi^{(+)}(\theta)\rangle_{12}$ and $|\Phi^{(+)}(\theta)\rangle_{34}$, where
$|\Phi^{(+)}(\theta)\rangle_{ij}=\cos\theta|1_H\rangle_i|1_H\rangle_j
+ \sin\theta|1_V\rangle_i|1_V\rangle_j$. Here, $|1_H\rangle_i$ refer to the state of a horizontally polarized single-photon occupying the spatial mode $i$. Quite clearly, the initial quantum state of all four photons can be written as,
\begin{eqnarray} \label{eq1}
|\Psi\rangle_{1234}&=&|\Phi^{(+)}(\theta)\rangle_{12}\otimes|\Phi^{(+)}(\theta)\rangle_{34}\nonumber\\
&=&\cos^2\theta|1_H,1_H,1_H,1_H\rangle_{1234}+\sin^2\theta|1_V,1_V,1_V,1_V\rangle_{1234}\nonumber\\
&+&\sin\theta\cos\theta\{|1_H,1_H,1_V,1_V\rangle_{1234}+|1_V,1_V,1_H,1_H\rangle_{1234}\}.
\end{eqnarray}

Next, photons 1-3 and 2-4 are made to interfere at the 50/50 beamsplitters BS1 and BS2, respectively, as shown in Fig.~\ref{fig}. Since independent photons are made to
interfere in this scheme, it is necessary that photon 1 and photon 3
(photon 2 and photon 4) must overlap at the beamsplitter BS1 (BS2)
perfectly. Experimentally, this means that the photon pair sources must operate in the pulse mode and the pulse width should be less than the coherence time of the individual photons.

At the beamsplitters BS1 and BS2, all the photons evolve according
to the well-known untiary transformation relation for a lossless
50/50 beamsplitter, for example, $|1_H\rangle_1 \rightarrow
\frac{1}{\sqrt{2}}(|1_H\rangle_{1'} + i|1_H\rangle_{3'})$. Since
photons in mode 1 and 3 (2 and 4) are made to interfere at BS1
(BS2), it is more convenient to rewrite the initial state as
\begin{eqnarray} \label{eq2}
|\Psi\rangle_{1234}
&=&\cos^2\theta|1_H,1_H\rangle_{13}|1_H,1_H\rangle_{24}+\sin^2\theta|1_V,1_V\rangle_{13}|1_V,1_V\rangle_{24}\nonumber\\
&+&\sin\theta\cos\theta\{|1_H,1_V\rangle_{13}|1_H,1_V\rangle_{24}+|1_V,1_H\rangle_{13}|1_V,1_H\rangle_{24}\}.
\end{eqnarray}

To calculate the quantum state after BS1 and BS2, we apply the beamsplitter unitary transformation to each term in eq.(\ref{eq2}). When BS1 and BS2 unitary transformations are applied to the first term in eq.~(\ref{eq2}), we find, 
\begin{eqnarray}
|1_H,1_H\rangle_{13}|1_H,1_H\rangle_{24} \rightarrow && -\frac{1}{2}(|2_H,0\rangle_{1'3'}+|0,2_H\rangle_{1'3'})\otimes
(|2_H,0\rangle_{2'4'}+|0,2_H\rangle_{2'4'})\nonumber\\
&&=-\frac{1}{2}(|2_H,2_H\rangle_{1'2'}+|2_H,2_H\rangle_{1'4'}+|2_H,2_H\rangle_{2'3'}+|2_H,2_H\rangle_{2'4'}),
\label{unitary1}
\end{eqnarray}
where the normalization factor for a number state is properly taken into account. Note that the unitary transformation described in eq.(\ref{unitary1}) is in fact due to Hong-Ou-Mandel interference as all the photons have the same polarization \cite{HOM}. Similarly, for the second term in eq.(\ref{eq2}), we have
\begin{eqnarray}
|1_V,1_V\rangle_{13}|1_V,1_V\rangle_{24} \rightarrow &&
-\frac{1}{2}(|2_V,0\rangle_{1'3'}+|0,2_V\rangle_{1'3'})\otimes
(|2_V,0\rangle_{2'4'}+|0,2_V\rangle_{2'4'}),\nonumber\\
&&=-\frac{1}{2}(|2_V,2_V\rangle_{1'2'}+|2_V,2_V\rangle_{1'4'}+|2_V,2_V\rangle_{2'3'}+|2_V,2_V\rangle_{2'4'}).
\label{unitary3}
\end{eqnarray}

Finally, for the third and the fourth terms in eq.(\ref{eq2}), we have,
\begin{eqnarray}
&&|1_H,1_V\rangle_{13}|1_H,1_V\rangle_{24}+|1_V,1_H\rangle_{13}|1_V,1_H\rangle_{24} \rightarrow\nonumber\\
&&~~~~~ -\frac{1}{2}\{(|1_H1_V,0\rangle_{1'3'}+|0,1_H1_V\rangle_{1'3'})\otimes(|1_H1_V,0\rangle_{2'4'}+|0,1_H1_V\rangle_{2'4'})\}\nonumber\\
&&~~~~~ +\frac{1}{2}\{(|1_H,1_V\rangle_{1'3'}-|1_V,1_H\rangle_{1'3'})\otimes(|1_H,1_V\rangle_{2'4'}-|1_V,1_H\rangle_{2'4'})\}\nonumber\\
&&~~~~~ =-\frac{1}{2}\{|1_H1_V,1_H1_V\rangle_{1'2'}+|1_H1_V,1_H1_V\rangle_{1'3'}
+|1_H1_V,1_H1_V\rangle_{2'3'}+|1_H1_V,1_H1_V\rangle_{3'4'}\}\nonumber\\
&&~~~~~~~~~ +|\Psi^{(-)}\rangle_{1'3'}\otimes|\Psi^{(-)}\rangle_{2'4'},\label{unitary4}
\end{eqnarray}
where $|\Psi^{(-)}\rangle_{ij}=\frac{1}{\sqrt{2}} (|1_H\rangle_i |1_V\rangle_j - |1_V\rangle_i |1_H\rangle_j)$ is the singlet Bell state. 

The quantum state of the photons after BS1 and BS2 unitary transformations, given the initial input state shown in eq.~(\ref{eq1}), can be found by substituting eq.~(\ref{unitary1})$\sim$eq.~(\ref{unitary4}) into eq.~(\ref{eq2}). The properly normalized quantum state of photons after BS1 and BS2 is, therefore, given as,
\begin{eqnarray}
|\Psi\rangle_{1'2'3'4'} &=& -\frac{1}{2}\{\cos^2\theta(|2_H,2_H\rangle_{1'2'}+|2_H,2_H\rangle_{1'4'}+|2_H,2_H\rangle_{2'3'}+|2_H,2_H\rangle_{2'4'})\nonumber\\
&&~~~~ +\sin^2\theta(|2_V,2_V\rangle_{1'2'}+|2_V,2_V\rangle_{1'4'}+|2_V,2_V\rangle_{2'3'}+|2_V,2_V\rangle_{2'4'})\nonumber\\
&&~~~~ +\sin\theta\cos\theta(|1_H1_V,1_H1_V\rangle_{1'2'}+|1_H1_V,1_H1_V\rangle_{1'3'}+|1_H1_V,1_H1_V\rangle_{2'3'}+|1_H1_V,1_H1_V\rangle_{3'4'})\nonumber\\
&&~~~~ -2\sin\theta\cos\theta|\Psi^{(-)}\rangle_{1'3'}\otimes|\Psi^{(-)}\rangle_{2'4'}\}.
\label{eq3}
\end{eqnarray}

%

Equation (\ref{eq3}) can be expressed in a more compact form by using the non-maximally entangled state notation we introduced earlier, i.e., $|\Phi^{(+)}(\theta)\rangle_{ij}=\cos\theta|1_H\rangle_i|1_H\rangle_j + \sin\theta|1_V\rangle_i|1_V\rangle_j$, where the subscripts $i$ and $j$ refer to the output modes of BS1 and BS2. Using this notation, the quantum state of photons after BS1 and BS2 can then be expressed in a simple form as,
\begin{eqnarray}
|\Psi\rangle_{1'2'3'4'} = -\frac{1}{4}\{|\Phi^{(+)}(\theta)\rangle^2_{1'2'}+|\Phi^{(+)}(\theta)\rangle^2_{1'4'}+|\Phi^{(+)}(\theta)\rangle^2_{2'3'}+|\Phi^{(+)}(\theta)\rangle^2_{3'4'} -2\sin2\theta|\Psi^{(-)}\rangle_{1'3'}\otimes|\Psi^{(-)}\rangle_{2'4'}\}.
\label{eq4}
\end{eqnarray}

The quantum state in eq.~(\ref{eq4}) exhibits interesting and important properties related to non-postselection entanglement concentration. First, the last term of eq.~(\ref{eq4}), $|\Psi^{(-)}\rangle_{1'3'}\otimes|\Psi^{(-)}\rangle_{2'4'}$, indicates that previously unrelated photons $1'$-$3'$ and $2'$-$4'$ are now entangled in the singlet Bell-states, i.e., entanglement swapping  has occurred.  The presence of this amplitude is the key element of the entanglement concentration. Second, the last term of eq.~(\ref{eq4}) is the only ampliude with exactly one photon occupying each output mode of BS1 and BS2.
This interesting feature of eq.~(\ref{eq4}) will be exploited to accomplish non-postselection entanglement concentration.

Now that we have calculated the quantum state of the photons at the output modes ($1'$, $2'$, $3'$, and $4'$) of BS1 and BS2, let us discuss how to accomplish non-postselection entanglement concentration. Quite clearly from eq.~(\ref{eq4}), the singlet Bell-state measurement (BSM) on the photon pair $1'$-$3'$ would project the photon pair $2'$-$4'$ to the maximally entangled singlet Bell-state, $|\Psi^{(-)}\rangle_{2'4'}$. So, if the singlet BSM can indeed be performed on the previously uncorrelated photons $1'$ and $3'$, its outcome can be used to signal successful preparation of the singlet Bell-state for photons $2'$-$4'$. 

The problem, however, is that the BSM is experimentally non-trivial and the complete BSM for photons require non-linear photon-photon interaction \cite{Kim5}. Since our scheme requires only the singlet BSM, one may try to use the linear optical singlet  BSM scheme: by interfering photons $1'$ and $3'$ at a 50/50 beamsplitter and detecting coincidence at the output ports of this beamsplitter \cite{Kim5,Braunstein}. This approach would not work in our scheme either because the coincidence event could equally likely come from the first four amplitudes in eq.~(\ref{eq4}). For example, the $|\Phi^{(+)}(\theta)\rangle^2_{1'2'}$ amplitude could cause the linear optical singlet BSM to trigger spuriously as there are two photons in mode $1'$. Since these spurious coincidence counts are indistinguishable from the desired ones, the conventional
linear optical singlet BSM method cannot be applied in our scheme to accomplish non-postselection entanglement concentration.

There exists, however, a simple projection measurement whose outcomes would deterministically signal successful preparation of the singlet Bell-state in modes $2'$-$4'$ provided that the input state is given exactly as in eq.~(\ref{eq1}). To see how this projection measurement works, let us first re-express eq.~(\ref{eq4}) as,
\begin{eqnarray} \label{eq5}
|\Psi\rangle_{1'2'3'4'} = -\frac{1}{4}\{ |\Phi^{(+)}(\theta)\rangle_{HOM}^2-\sqrt{2}\sin2\theta (|H_{1'}V_{3'}\rangle-|V_{1'}H_{3'}\rangle)
{|\Psi^{(-)}\rangle}_{2'4'}\},
\end{eqnarray}
where $|\Phi^{(+)}(\theta)\rangle_{HOM}^2\equiv{|\Phi^{(+)}(\theta)\rangle}_{1'2'}^2+{|\Phi^{(+)}(\theta)\rangle}_{1'4'}^2+{|\Phi^{(+)}(\theta)\rangle}_{2'3'}^2+{|\Phi^{(+)}(\theta)\rangle}_{3'4'}^2$. Here we have expressed the singlet Bell-state $|\Psi^{(-)}\rangle_{1'3'}$ in the H-V basis but it makes no difference to choose another basis, such as, rotated-linear or circular basis. 

If we now make the following joint polarization projection measurement (i.e., coincidence measurement) on photons $1'$ and $3'$, photons $2'$ and $4'$ will be found in the singlet Bell-state:
\begin{equation}\label{eq6}
\langle H_{1'}V_{3'}|\Psi\rangle_{1'2'3'4'}=\langle
V_{1'}H_{3'}|\Psi\rangle_{1'2'3'4'}=|\Psi^{(-)}\rangle_{2'4'}.
\end{equation}
The non-postselection nature of this scheme comes from the fact that the projection measurement described in eq.~(\ref{eq6}) only affects the desired amplitudes in eq.~(\ref{eq5}). That is, there are no spurious measurement outcomes as long as the input state (two identical pairs of non-maximally entangled photons) is provided as requested in eq.~(\ref{eq1}).

The experimental scheme to implement this projection measurement is shown in Fig.~\ref{fig}. Polarizing beamsplitters PBS1 and PBS2 are placed (in the H-V basis) at the output ports of BS1. At each output port of PBS1 and PBS2, a single-photon detector is installed and coincidence events between detectors $\{D_{1'H}, D_{3'V}\}$ and $\{D_{1'V}, D_{3'H}\}$ are recorded. Since the joint projection measurement is for the spatial modes $1'$ and $3'$, $|\Phi^{(+)}(\theta)\rangle_{HOM}^2$ amplitude in eq.~(\ref{eq5}) contributes no unwanted spurious coincidence events. Therefore, a coincidence event between the detectors  $\{D_{1'H}, D_{3'V}\}$ or $\{D_{1'V}, D_{3'H}\}$ signals
a successful projection or preparation of photons $2'$ and $4'$ into
the singlet Bell state, deterministically.

\section{Discussion}

The present non-postselection entanglement concentration scheme has a few interesting features that we wish to point out. First, the joint projection measurement using PBS1 and PBS2 need not be done in the H-V basis. A rotated-linear basis, such as, $|D\rangle$-$|A\rangle$ basis, or a circular basis $|R\rangle$-$|L\rangle$ would work equally well. Here $|D\rangle = (|H\rangle+|V\rangle)/\sqrt{2}$, $|A\rangle = (|H\rangle-|V\rangle)/\sqrt{2}$, $|L\rangle = (|H\rangle+i|V\rangle)/\sqrt{2}$, and $|R\rangle = (|H\rangle-i|V\rangle)/\sqrt{2}$. This is due to the fact that the singlet Bell-state is rotationally invariant so that it retains its form in any basis, i.e., $|\Psi^{(-)}\rangle_{1'3'} =
\frac{1}{\sqrt{2}}(|D\rangle_{1'}|A\rangle_{3'}-|A\rangle_{1'}|D\rangle_{3'}) =
\frac{1}{\sqrt{2}}(|R\rangle_{1'}|L\rangle_{3'}-|L\rangle_{1'}|R\rangle_{3'})$. Therefore, instead of inserting PBS1 and PBS2 as shown in Fig.~\ref{fig}, one could use any kind of polarization projection devices as long as the coincidence circuits detect orthogonally
polarized photons in modes $1'$ and $3'$, e.g., $|D\rangle_{1'}$ and $|A\rangle_{3'}$ or
$|R\rangle_{1'}$ and $|L\rangle_{3'}$. And, regardless of the measurement choice, the output state remains to be the singlet Bell-state. 

The second interesting feature of the present scheme is that it can be applied to any types of non-maximally entangled two-photon polarization states as long as two initial pairs of photons are identically prepared. This can be shown by substituting the input state in eq.~(\ref{eq1}) with one of the other non-maximally entangled states. After some straightforward calculations, it can be shown that the output states are given as,
\begin{eqnarray}
|\Psi^{(\pm)}(\theta)\rangle_{12}|\Psi^{(\pm)}(\theta)\rangle_{34} \rightarrow && -\frac{1}{4}\{ |\Psi^{(\pm)}(\theta)\rangle_{HOM}^2 \mp 2\sin2\theta{|\Psi^{(-)}\rangle}_{1'3'}{|\Psi^{(-)}\rangle}_{2'4'}\},\nonumber\\
|\Phi^{(-)}(\theta)\rangle_{12}|\Phi^{(-)}(\theta)\rangle_{34}\rightarrow &&-\frac{1}{4}\{|\Phi^{(-)}(\theta)\rangle_{HOM}^2 + 2\sin2\theta{|\Psi^{(-)}\rangle}_{1'3'}{|\Psi^{(-)}\rangle}_{2'4'}\},\nonumber
\end{eqnarray}
where $|\Psi^{(\pm)}(\theta)\rangle_{ij}=\cos\theta|H\rangle_i |V\rangle_j \pm\sin\theta|V\rangle_i |H\rangle_j$, $|\Phi^{(\pm)}(\theta)\rangle_{ij}=\cos\theta|H\rangle_i |H\rangle_j \pm\sin\theta|V\rangle_i |V\rangle_j$,  $|\Psi^{(\pm)}(\theta)\rangle_{HOM}^2 \equiv {|\Psi^{(\pm)}(\theta)\rangle}_{1'2'}^2
+{|\Psi^{(\pm)}(\theta)\rangle}_{1'4'}^2+{|\Psi^{(\pm)}(\theta)\rangle}_{2'3'}^2
+{|\Psi^{(\pm)}(\theta)\rangle}_{3'4'}^2 $, and
$|\Phi^{(-)}(\theta)\rangle_{HOM}^2={|\Phi^{(-)}(\theta)\rangle}_{1'2'}^2+{|\Phi^{(-)}(\theta)\rangle}_{1'4'}^2+{|\Phi^{(-)}(\theta)\rangle}_{2'3'}^2
+{|\Phi^{(-)}(\theta)\rangle}_{3'4'}^2$. 

Since the photon pairs $1'$-$3'$ and $2'$-$4'$ are in the same singlet Bell-state regardless of the input state, we can apply the same projection measurement on photons $1'$ and $3'$ as before and the output photon pair in modes $2'$ and $4'$ will always be found in the singlet Bell-state. 

For the present non-postselection entanglement concentration scheme to work as described, the initial state must be provided as requested in eq.~(\ref{eq1}). In other words, the initial non-maximally entangled photon pairs can be supplied probabilistically (this will only reduce the success probability), but there should not be any double-pair events from any of two the sources. Should there be two pairs of non-maximally entangled photons either in modes $1$-$2$ or $3$-$4$, the projection measurement outcomes would now contain spurious coincidence events that come from these double-pair events and the output photon pair in modes $2'$-$4'$ would no longer be in the singlet Bell-state, similarly to the result of Ref.~\cite{Wang}. 

Therefore, if this scheme is implemented using spontaneous parametric down-conversion sources \cite{kwiat1,Kim4}, the non-postselection nature of the scheme cannot be utilized due to the inherent multi-pair events probabilities of SPDC \cite{Kok}. The proposed non-postselection entanglement concentration scheme, however, could be implemented in full using recently reported semiconductor quantum-dot entangled photon sources in which the multi-pair events could be suppressed \cite{akopian}.

\section{Conclusion}

We have discussed a practical scheme for non-postselection entanglement concentration using 50/50 beamsplitters, polarizing beamsplitters and conventional photon detectors. The scheme has the success probability less than 25\% given two identical pairs of non-maximally entangled photons but the successful projection measurement deterministically prepares the output photon pair in the singlet Bell-state. In addition, unlike other entanglement concentration schemes discussed previously, the present scheme works for all non-maximally entangled two-photon polarization entangled states as the input.  A possible drawback of this scheme is that, due to the specific nature of quantum interference and the projection measurement used in this scheme, it may not be well suited as a long-distance entanglement repeater. We, however, believe that this scheme will have applications in experimental implementation of various quantum information protocols, which require high-quality deterministic sources of two-photon Bell-states.

\section*{Acknowledgement}

This work was supported, in part, by the Korea Research Foundation (R08-2004-000-10018-0 and KRF-2005-015-C00116), the Korea Science and Engineering Foundation (R01-2006-000-10354-0), and POSTECH.


\end{document}